\documentclass[a4paper,11pt]{article}
\usepackage{graphics}
\usepackage{jheppub}
\usepackage{epsfig}
\usepackage{float}
\usepackage{slashed}
\usepackage{soul}
\usepackage{cancel}
\usepackage{color}
\definecolor{ourcolor}{rgb}{0.7, 0.25, 0.05}
\expandafter\ifx\csname package@font\endcsname\relax\else
 \expandafter\expandafter
 \expandafter\usepackage
 \expandafter\expandafter
 \expandafter{\csname package@font\endcsname}%
\fi

\title{\color{ourcolor} Lorentz Invariance Violation and IceCube Neutrino Events}

\author[a,b]{Gaurav Tomar,}
\author[a]{Subhendra Mohanty}
\author[c]{and Sandip Pakvasa}
\affiliation[a]{Physical Research Laboratory, Ahmedabad 380009, India.}
\affiliation[b]{Indian Institute of Technology, Gandhinagar 382424, India.}
\affiliation[c]{Department of Physics \& Astronomy, University of Hawaii, Honolulu, HI 96822, USA.}

\emailAdd{tomar@prl.res.in}
\emailAdd{mohanty@prl.res.in}
\emailAdd{pakvasa@phys.hawaii.edu}

\abstract{ The IceCube neutrino spectrum shows a flux which falls of as $E^{-2}$ for sub PeV energies but there are  no 
neutrino events observed  above $\sim 3$ PeV. In particular  the Glashow resonance expected at 6.3 PeV is not 
seen. We examine a Planck scale Lorentz violation as a mechanism for explaining the cutoff of observed neutrino energies around a few PeV. By choosing the one free parameter the cutoff in neutrino energy can be chosen to be between 2 and 6.3 PeV. We assume that neutrinos (antineutrinos) have a dispersion relation $E^2=p^2 - (\xi_3/M_{Pl})~p^3$, and find that both $\pi^+$ and $\pi^-$ decays are suppressed at neutrino energies of order of few PeV. We find that the $\mu^-$ decay being a two-neutrino process is enhanced, whereas $\mu^+$ decay is suppressed. The $K^+\rightarrow \pi^0 e^+ \nu_e$ is also suppressed with a cutoff neutrino energy of same order of magnitude, whereas $K^-\rightarrow \pi^0 e^- \bar \nu_e$ is enhanced. The $n \rightarrow p^+ e^- \bar \nu_e$ decay is suppressed (while the $\bar n \rightarrow p^- e^+ \nu_e$ is enhanced). This means that the $\bar \nu_e$ expected from $n$ decay arising from $p+\gamma \rightarrow \Delta \rightarrow \pi^+ n$ reaction will not be seen. This can explain the lack of Glashow resonance events at IceCube. If no Glashow resonance events are seen in the future then the Lorentz violation can be a viable explanation for the IceCube observations at PeV energies.}
\begin{document}
\maketitle
\flushbottom

\section{Introduction}
IceCube collaboration has observed the neutrinos of very high energy going to beyond $2.6$ PeV 
\cite{Aartsen:2013bka, Aartsen:2013jdh, Aartsen:2014gkd,icecube}. 
The IceCube data in the energy range 60 TeV to $\sim 3$ PeV is consistent with $E_\nu^{-2}$ neutrino spectrum following $E^2_\nu dN_\nu/dE_\nu \simeq 1.2\times 10^{-8}~\mbox{GeV}\rm cm^{-2} s^{-1} sr^{-1}$ \cite{Aartsen:2013jdh, Aartsen:2014gkd}. A neutrino spectrum sharper than $E^{-2.3}$ does not give a good fit to the 
data \cite{Aartsen:2014gkd}. There are no neutrino events observed above $\sim 3$ PeV.\\
In particular, there is no evidence of  Glashow resonance \cite{Glashow:1960zz}, $\bar \nu_e + e^- \rightarrow W^- \rightarrow $ shower, which is expected at $E=6.3$ PeV. Glashow resonance gives rise to an enhanced cross-section for $\bar \nu_e$ at
resonance energy $E = M^2_W/2 m_e = 6.3$ PeV, which increases the detection rate of $\nu_e + \bar \nu_e$ by a 
factor of $\sim 10$ \cite{Aartsen:2013jdh}. This implies that at least three events should have been observed at Glashow 
resonance, but none were.\\ 
The Glashow resonance gives rise to multiple energy peaks at different energies
\cite{Kistler:2013my}. The first one is at $6.3$ PeV and others lie at the $E_{vis} = E - E_X$, where 
$E_X$ is the energy in the $W$ decay, which does not contribute to the visible shower \cite{Anchordoqui:2014hua}. The 
decay of $W$ into hadrons goes as $W\rightarrow \bar q q$, giving rise to a peak at $6.3$ PeV, while decay into leptons
goes as $W \rightarrow \bar \nu l$, which means $W$ boson will lose half of its energy and so a second peak at 3.2 PeV
is expected. In case of $\tau$ lepton in the final state, a further decay takes place producing a neutrino and thus a third peak at 1.6 PeV. The events observed by IceCube \cite{Aartsen:2013bka, Aartsen:2013jdh, Aartsen:2014gkd,icecube} between 1 PeV to $\sim 3$ PeV range may be associated with the second (leptonic decay of $W$) and third peak ($\tau$ decay), but non-appearance of Glashow resonance hadronic shower from $W\rightarrow \bar q q$ at $6.3$ PeV (dominant peak) makes this idea less attractive. The non observation of the expected signature of Glashow resonance in IceCube data indicates a cutoff of neutrino energies between 2-6.3 PeV \cite{Anchordoqui:2014hua,Learned:2014vya}.\\
In this paper, we propose a mechanism which can explain why neutrinos above a certain energy may be suppressed in the 
astrophysical production processes like $\pi \rightarrow \mu \nu_\mu,~K \rightarrow \mu \nu_\mu$ etc. We assume that Lorentz violating higher dimensional operators \cite{Myers:2003fd,Diaz:2013wia}
give rise to a modified dispersion relation for the neutrinos (antineutrinos) of the form $E^2= p^2 + m_\nu^2 - (\xi_n/M_{pl}^{n-2})~p^n$ with $n>2$. Depending on the sign of $\xi_n$, the neutrinos (antineutrinos) can be either superluminal $(\xi_n<0)$ or subluminal $(\xi_n>0)$. For the superluminal case, it has been shown \cite{Mohanty:2011aa,  Stecker:2014oxa} that the presence of the extra terms in the dispersion results in a suppression of $\pi$ and $K$ decay widths. The phase space suppression for both the subluminal and superluminal dispersions for meson decay and the Cerenkov process $\nu\rightarrow \nu e^+ e^-$ has been noticed in \cite{Kostelecky:2011gq,Diaz:2013wia,Diaz:2014yva,Stecker:2014xja,Maccione:2011fr} with limits on Lorentz violation parameters from IceCube events. A comprehensive listing of Lorentz and CPT violating operators and their experimental constraints is given in \cite{Kostelecky:2008ts}. In this paper, we calculate the $\pi,K,\mu$ and $n$ decay processes in a fixed frame (the frame chosen being the one in which the CMBR is isotropic; although the Earth moves at a speed $v_{Earth}\sim 300$ km/sec with respect to the CMBR, the Lorentz correction to the neutrino energy is small as $\beta_{Earth}\sim 10^{-3}$), 
where the neutrinos (antineutrinos) dispersion relation is $E^2= p^2 + m_\nu^2 - (\xi_3/M_{pl})~p^3$ 
\cite{Bolokhov:2007yc,Myers:2003fd,Jacobson:2003bn,Jacobson:2005bg}. We will have $\xi_3>0$ for neutrinos and $\xi_3<0$ for  antineutrinos. In the $\pi^+$ decay, we find that the $\overline{|M|^2}$ is suppressed at neutrino energy $E_\nu$, where $m^2_\pi-m^2_\mu \simeq (\xi_3 /M_{pl})~p_\nu^3$. This implies that for the leading order Planck suppression $(n=3)$ taking $\xi_3 \sim 0.05$, the $\pi^+$ decay is suppressed at $E_\nu \sim 1.3$ PeV. Similarly $K^+$ decay will be cutoff at $E_\nu\sim 2$ PeV with $m^2_K-m^2_\mu\sim (\xi_3/M_{pl})p^3$ and neutron decay will be cutoff for $p$, where $(m_n-m_p)^2\sim (\xi_3/M_{pl})p^3$, which is lower than the Glashow resonance energy. For the $\pi^-$ decay the $\overline{|M|^2}$ is enhanced but the phase space is suppressed and therefor 
$\pi^-\rightarrow \mu^-\nu_\mu$ is also suppressed. In the case of $\mu^-\rightarrow e^-\bar \nu_e \nu_\mu$ decay, 
$\overline{|M|^2}$ is enhanced whereas the phase space suppression is not significant, so the $\mu^-$ decay rate is enhanced 
(while $\mu^+\rightarrow e^+ \nu_e \bar \nu_\mu$ decay rate is suppressed).
This enhancement is significant at $\mu^-$ energies $\sim 2$ PeV but since the primary source of $\mu^-$ is $\pi^-$ 
decay which is already cutoff, there will be no observable effect of this enhancement in the neutrino spectrum seen 
at IceCube. Neutrinos from $K^- \rightarrow \mu^- \bar \nu_\mu$ and $K^+ \rightarrow \mu^+ \nu_\mu$ decays will be cutoff at slightly higher energies. Radiative $\pi^\pm$ decay with a single neutrino in the outgoing state are also suppressed. The three body kaon decay rate are determined by the $\xi_3$ dependence of $\overline{|M|^2}$ and we find that $K^+ \rightarrow \pi^0 \mu^+ \nu_\mu$ decay is suppressed but $K^- \rightarrow \pi^0 \mu^- \bar \nu_\mu$ decay is enhanced.
Neutron beta decay $n \rightarrow p^+ e^-\bar \nu_e$ gets suppressed in the same way as $\mu^+$ decay.
If the source of $\bar \nu_e$ is neutron beta-decay \cite{Sahu:2014fua} then the mechanism proposed in this paper can be used for explaining the absence of Glashow resonance \cite{Glashow:1960zz} at IceCube. The value of $(\xi_3/M_{pl})\sim 0.05~M^{-1}_{pl}$ used in this paper to explain the cutoff in PeV neutrinos is much smaller than the bound on the dimension-five coefficient, $(a^{(5)}_{\tiny \mbox{of}})_{00}~\textless~3.5 \times 10^{-10}~\mbox{GeV}^{-1}$ from SN1987A dispersion \cite{Kostelecky:2011gq}.\\
The rest of the paper is organized as follows. In section \ref{sec:twbd}, we calculate the leptonic decay widths of pions and 
kaons using modified dispersion relation of neutrino and compare them with their standard model counterparts. 
In section \ref{sec:thbd} we study $\mu^-\rightarrow e^-\bar \nu_e \nu_\mu$, $K^+\rightarrow \pi^0 e^+ \nu_e$ and 
$n\rightarrow p^+ e^- \bar \nu_e$ processes with 
modified neutrino dispersion. We give our conclusion in section \ref{sec:conclusion}.
\section{Two body decays}\label{sec:twbd}
\subsection{Neutrino velocity with modified dispersion}
To calculate the decay widths of pion, kaon and muon, we use the following dispersion relation,
\begin{equation}
 E^2 = p^2 + m^2_\nu -\frac{\xi_n}{M_{pl}^{n-2}}~p^n
 \label{dr}
\end{equation}
which is motivated by Lorentz violating higher dimensional operators \cite{Myers:2003fd,Diaz:2013wia}. We will take $\xi_n>0$ for neutrinos and $\xi_n<0$ for antineutrinos. We use this modified dispersion relation to get the neutrino (antineutrino) velocity, which becomes
\begin{equation}
 v = \frac{\partial E}{\partial p} = 1 - \frac{n-1}{2}~\frac{\xi_n}{M_{pl}^{n-2}} ~p^{n-2}
 \label{nu:vel}
\end{equation}
This is clear from eq.(\ref{nu:vel}) that we have a subluminal neutrinos and superluminal antineutrinos.
In this paper, we will consider the leading order Planck suppressed dispersion relation 
$E^2 = p^2 + m^2_\nu -(\xi_3/M_{pl})~p^3$ to compute the primary decay processes which produce neutrinos and antineutrinos. 
In appendix.\ref{ap:dspr}, we obtained modified dispersion relations for neutrinos and antineutrinos using dimension 5 operator.
\subsection{$\pi^+\rightarrow \mu^+ \nu_\mu$}
\begin{figure}[t!]
\vspace*{10 mm}
\begin{center}
\includegraphics[scale=1.10,angle=0]{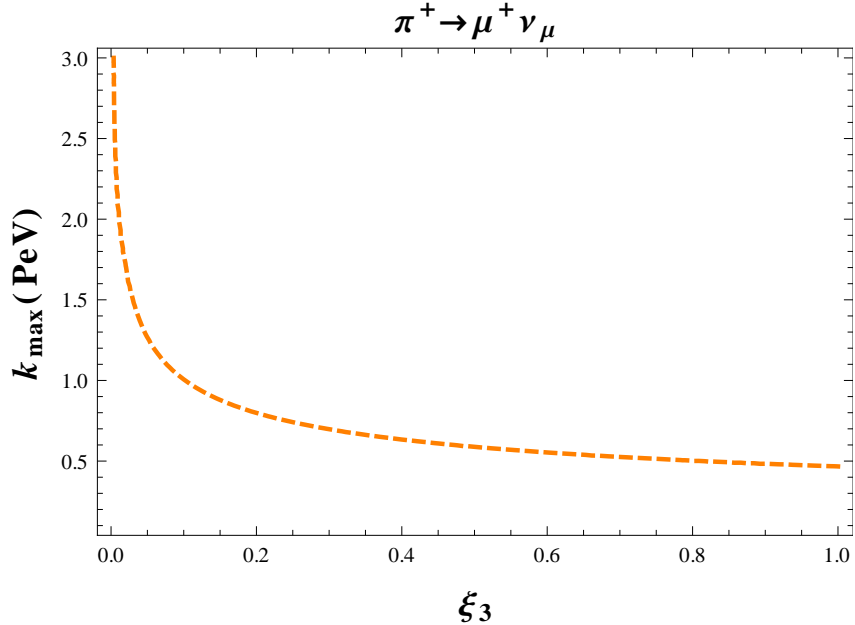}
\vspace*{3mm}
\caption{The maximum neutrino energy, $k_{max}$ as a function of Lorentz  invariance violation parameter $\xi_3$.}
\label{fig:km}
\end{center}
\end{figure}

We calculate the pion decay width using the modified dispersion relation for neutrino by taking $n=3$ case.
The amplitude calculation of pion decay process $\pi^+(q)\to\mu^+(p) \nu_\mu(k)$ gives,
\begin{figure}[t!]
\vspace*{10 mm}
\begin{center}
\includegraphics[scale=1.20,angle=0]{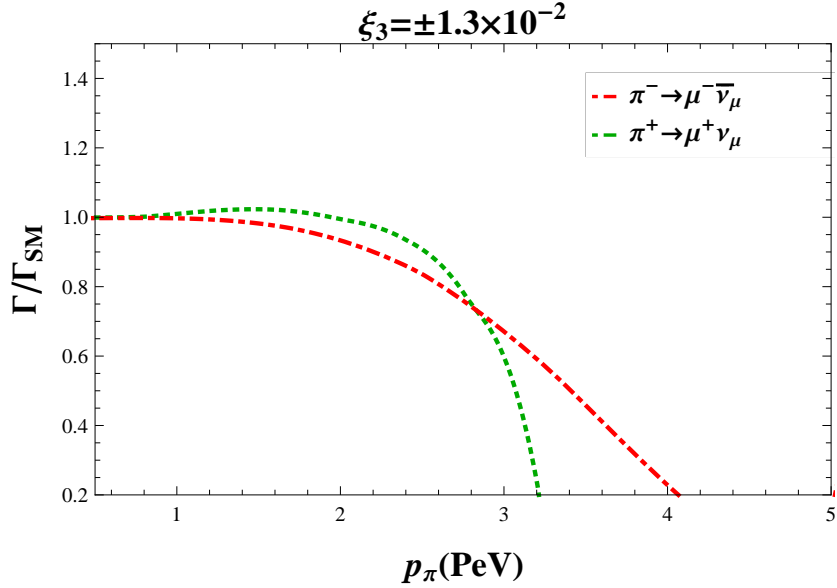}
\vspace*{3mm}
\caption{The ratio $\Gamma/\Gamma_{SM}$ for $\pi^+ \rightarrow \mu^+ \nu_\mu$ and $\pi^- \rightarrow \mu^- \bar \nu_\mu$ processes in Lorentz invariance violating framework to its standard model prediction for superluminal $\bar \nu_\mu~(\xi_3<0)$ and subluminal $\nu_\mu~(\xi_3>0)$ final states as a function of pion momentum $p_\pi$. We considered $\xi_3=\pm 1.3\times 10^{-2}$ for corresponding processes.} 
\label{fig:pi}
\end{center}
\end{figure} 
\begin{figure}[t!]
\vspace*{10 mm}
\begin{center}
\includegraphics[scale=1.20,angle=0]{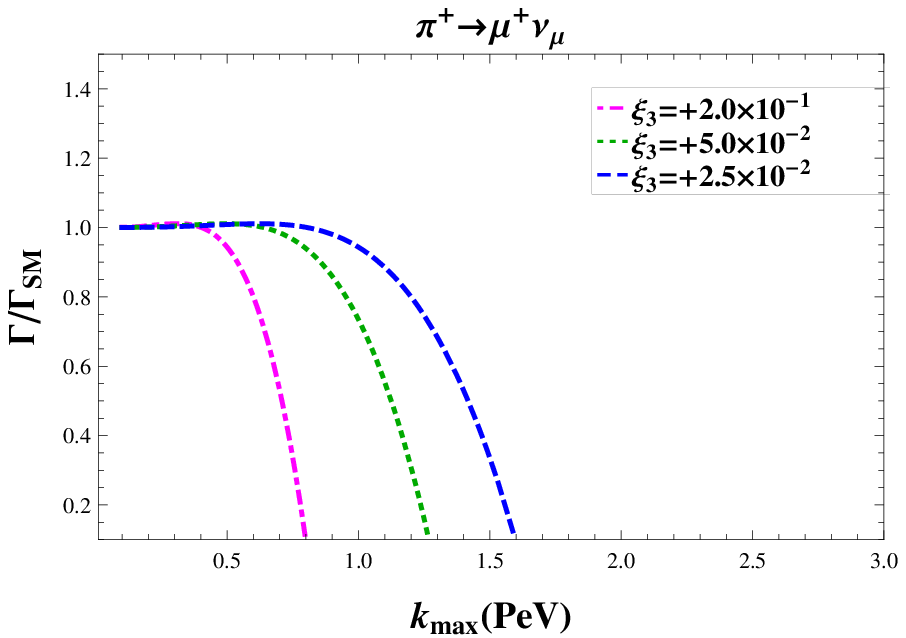}
\vspace*{3mm}
\caption{The ratio $\Gamma/\Gamma_{SM}$ of $\pi^+\rightarrow \mu^+ \nu_\mu$ process in Lorentz invariance violating 
framework to its standard model prediction for subluminal neutrino $(\xi_3>0)$ as a function of neutrino energy $k_{max}$ with different values of $\xi_3$.}
\label{fig:nu}
\end{center}
\end{figure} 
\begin{equation}
 M = f_{\pi} V_{ud}~q^\mu \bar u(k) \frac{G_F}{\sqrt{2}} \gamma_\mu (1-\gamma_5) v(p)
\end{equation}
where $f_{\pi}\equiv f(m^2_\pi)$ is a constant factor and $V_{ud}$ is the CKM matrix element. The spin averaged amplitude squared is,
\begin{equation}\label{amps}
 \overline{|M|^2}= 2G_F^2f_\pi^2 |V_{ud}|^2 m_\mu^2 F(k)\left[m_\pi^2-m_\mu^2 - \xi^\prime_3 k^3
 \left(\frac{m_\pi^2}{m_\mu^2}+2\right)\right]
\end{equation}
where $\xi^\prime_3\equiv \xi_3/M_{pl}$ and the $F(k)$ factor comes from the modified spinor relation of neutrino, 
as described in eq.(\ref{sra}). The decay width of pion is then given by,
\begin{align}
 \Gamma=&\frac{G_F^2f_\pi^2 |V_{ud}|^2 m_\mu^2 F(k)}{8\pi E_\pi}\int\,
	\frac{k^2\,dk\,d\cos\theta}{E_\nu \sqrt{|\vec{q}-\vec{k}|^2+m_\mu^2}}
	\delta(E_{\nu_\mu}-E_\pi+\sqrt{|\vec{q}-\vec{k}|^2+m_\mu^2})\nonumber\\
	&\times \left[m_\pi^2-m_\mu^2-  \xi^\prime_3 k^3\left(\frac{m_\pi^2}{m_\mu^2}+2\right)\right]
\label{pi-rate-f}
\end{align}
after using $E_{\nu_\mu}=F(k)k$, and writing 
$|\vec{p}| = |\vec{q}-\vec{k}|^2=k^2+q^2-2kq\cos\theta$, our expression of eq.(\ref{pi-rate-f}) takes the following form
\begin{align}
 \Gamma=&\frac{G_F^2f_\pi^2 |V_{ud}|^2 m_\mu^2}{8\pi E_\pi}\int\,
	\frac{k\,dk\,d\cos\theta}{\sqrt{|\vec{q}-\vec{k}|^2+m_\mu^2}}
	\delta(E_{\nu_\mu} -E_\pi+\sqrt{|\vec{q}-\vec{k}|^2+m_\mu^2})\nonumber\\
	&\times \left[m_\pi^2-m_\mu^2 -\xi^\prime_3 k^3\left(\frac{m_\pi^2}{m_\mu^2}+2\right)\right]
\label{pi-rate-r}
\end{align}
from the argument of the delta function in eq.(\ref{pi-rate-r}), we have
\begin{equation}
 \sqrt{|\vec{q}-\vec{k}|^2+m_\mu^2} = E_{\pi} -E_{\nu_\mu}
\end{equation}
which gives,
\begin{equation}
 \cos\theta= \frac{\left(m_\mu^2-m_\pi^2+2E_\pi k -  E_\pi k^2  \xi^\prime_3 + k^3  \xi^\prime_3\right)}{2kq} .
\label{cos}
\end{equation}
We reduce the $\delta$ function in $E_{\nu_\mu}$ to a $\delta$ function in $\cos\theta$ by taking,
\begin{equation}
 \left|\frac{d}{d\cos\theta}(E_{\nu_\mu} -E_\pi + \sqrt{|\vec{q}-\vec{k}|^2+m_\mu^2})\right|=
\frac{kq}{\sqrt{k^2+q^2-2kq\cos\theta+m_\mu^2}}
\end{equation}
and substituting in eq.(\ref{pi-rate-r}). We get the pion decay width,
\begin{equation}
 \Gamma=\frac{G_F^2f_\pi^2 |V_{ud}|^2 m_\mu^2}{8\pi E_\pi}\int\,\frac{dk}{q}
        \left[m_\pi^2-m_\mu^2 - \xi^\prime_3 k^3\left(\frac{m_\pi^2}{m_\mu^2}+2\right)\right].
\label{pidw}        
\end{equation}
We solve the integration in the limits of $k$, which are fixed by taking $\cos\theta=\pm 1$ in eq.(\ref{cos}),
\begin{equation}
 k_{max}= \frac{m_\pi^2-m_\mu^2 +  \xi^\prime_3 k_{max}^2(E_\pi-k_{max})}{2(E_\pi-q)}
 \label{kmax}
\end{equation}
\begin{equation}
 k_{min}= \frac{m_\pi^2-m_\mu^2 + \xi^\prime_3 k_{min}^2(E_\pi-k_{min})}{2(E_\pi+q)}
 \label{kmin}
\end{equation}
solving these equations numerically, we get the allowed limits of neutrino momentum. 
We solve eq.(\ref{pidw}) and then compare our result with the standard model result of pion decay in a moving frame, which is
\begin{equation}
 \Gamma_{SM}(\pi\to \mu \nu)=\frac{G^2_F f^2_{\pi} |V_{ud}|^2 m^2_{\mu}m_\pi^2}{8\pi E_\pi}\left(1-\frac{m_\mu^2}{m_\pi^2}\right)^2.
\end{equation}
We compute the pion decay rate numerically for superluminal $\bar \nu_e~(\xi_3<0)$ and subluminal 
$\nu_e~(\xi_3>0)$ final states and obtain the following :
\begin{itemize}
 \item For subluminal neutrino final state $(\xi_3>0)$, the 
allowed phase space (eq.\ref{kmax}-eq.\ref{kmin}) goes up but the $\overline{|M|^2}$ (eq.\ref{amps}) is suppressed.
There is a net suppression in $\Gamma(\pi^+\rightarrow \mu^+ \nu_\mu)$ as shown in Fig.(\ref{fig:pi}) for $\xi_3 = 1.3\times 10^{-2}$.
 \item For superluminal antineutrino final state $(\xi_3<0)$, the phase space (eq.\ref{kmax}-eq.\ref{kmin}) is 
suppressed but the $\overline{|M|^2}$ is enhanced. The net effect however is a suppression in the $\Gamma(\pi^-\rightarrow \mu^- \bar \nu_\mu)$ for this case also \cite{Mohanty:2011aa}, as shown in Fig.(\ref{fig:pi}) for $\xi_3 = -1.3\times 10^{-2}$.
\end{itemize}
In Fig.(\ref{fig:nu}), for the process $\pi^+\rightarrow \mu^+ \nu_\mu$, we show the maximum neutrino energy for 
different values of $\xi_3$ using the solution for $q$ in terms of $k_{max}$ and $k_{min}$ from eq.(\ref{kmax}-\ref{kmin}) in eq.(\ref{pidw}). We see that for $\xi_3=5.0\times 10^{-2}$, the neutrino spectrum cutoff at $k_{max}=1.3$ PeV.
The upper limit of observed neutrino energy provides bound on the Lorentz invariance violation parameter $\xi_3$. In 
Fig.(\ref{fig:km}), we show the maximum neutrino energy $k_{max}$, as a function of Lorentz invariance violation 
parameter $\xi_3$. This is clear from Fig.(\ref{fig:km}) that $k_{max}$ goes down as $\xi_3$ increases. 
%
%
\begin{figure}[h!]
\vspace*{10 mm}
\begin{center}
\includegraphics[scale=1.20,angle=0]{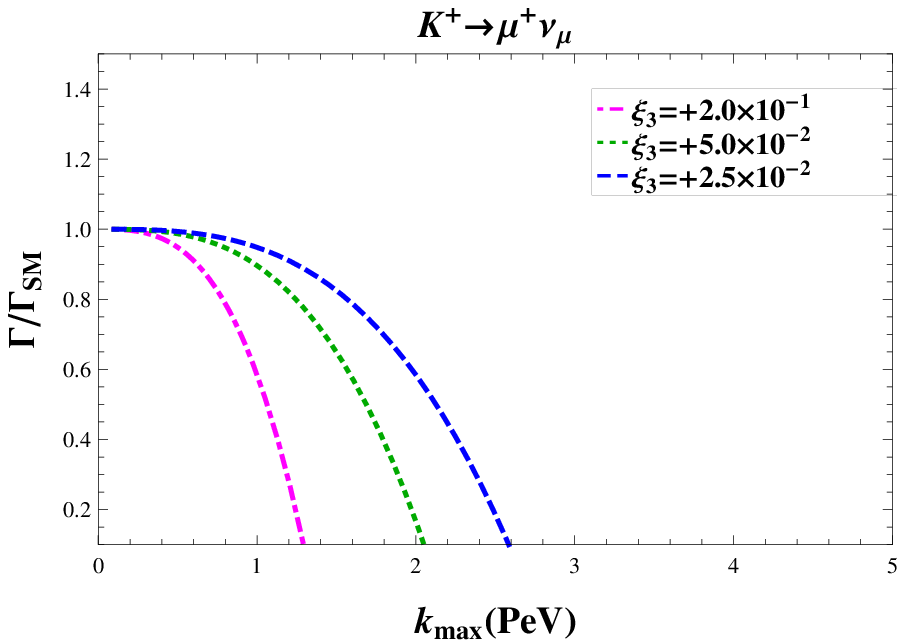}
\vspace*{3mm}
\caption{The ratio $\Gamma/\Gamma_{SM}$ of $	K^+\rightarrow \mu^+ \nu_\mu$ process in Lorentz invariance violating 
framework to its standard model prediction for subluminal neutrino $(\xi_3>0)$ as a function of neutrino energy $k_{max}$ with 
different values of $\xi_3$.}
\label{fig:nu_k}
\end{center}
\end{figure} 
%
\subsection{$\boldmath {K^+\rightarrow \mu^+ \nu_\mu}$}
In the similar way like pion decay, we calculate the kaon decay width for the process $K^+(q)\to\mu^+(p)\nu_\mu(k)$, 
using the modified dispersion relation for neutrinos by taking $n=3$ case. 
We get the kaon decay width,
\begin{equation}
 \Gamma=\frac{G_F^2f_K^2 |V_{us}|^2 m_\mu^2}{8\pi E_K}\int\,\frac{dk}{q}
        \left[m_K^2-m_\mu^2 - \xi^\prime_3 k^3\left(\frac{m_K^2}{m_\mu^2}+2\right)\right].
\label{kdw}        
\end{equation}
In the same way like pion, we solve the integration in the limits of $k$ by taking $\cos\theta=\pm 1$ which gives,
\begin{equation}
 k_{max}= \frac{m_K^2-m_\mu^2 +  \xi^\prime_3 k_{max}^2(E_K-k_{max})}{2(E_K-q)}
 \label{kmax_k}
\end{equation}
\begin{equation}
 k_{min}= \frac{m_K^2-m_\mu^2 + \xi^\prime_3 k_{min}^2(E_K-k_{min})}{2(E_K+q)}
 \label{kmin_k}
\end{equation}
solving these equations numerically, we get the allowed limits of neutrino momentum. 
We solve eq.(\ref{kdw}) and then compare our result with the standard model result of kaon decay in a moving frame, which is
\begin{equation}
 \Gamma_{SM}(K\to \mu \nu)=\frac{G^2_F f^2_{K} |V_{us}|^2 m^2_{\mu}m_K^2}{8\pi E_K}\left(1-\frac{m_\mu^2}{m_K^2}\right)^2.
\end{equation}
In Fig.(\ref{fig:nu_k}), we show the maximum neutrino energy for different values of $\xi_3$ using the solution for $q$ in
terms of $k_{max}$ and $k_{min}$ from eq.(\ref{kmax_k}-\ref{kmin_k}) in eq.(\ref{kdw}). We see that 
for $\xi_3=5.0\times 10^{-2}$ the neutrino spectrum cutoff at $k_{max}=2$ PeV.
\section{Three body decays}\label{sec:thbd}
\subsection{$\boldmath {\mu^-\rightarrow e^- \bar \nu_e \nu_\mu}$}
We compute the muon decay width with subluminal neutrino and superluminal anti-neutrino in the final state, 
assuming the dispersion relation for the neutrino (antineutrino), $E^2_\nu = k^2 - \xi^\prime_3 k^3$, where $\xi_3 >0$ and $\xi_3 <0$ correspond to subluminal neutrino and superluminal antineutrino respectively. We assume identical $\xi_3$ for all the species of $\nu~(\mbox{and}~\bar\nu)$ to avoid an extra source for neutrino oscillations which is not observed \cite{Giudice:2011mm,Maccione:2011fr}.
The amplitude for the process $\mu^-(p)\rightarrow e^-(k^\prime) \bar \nu_e(k) \nu_\mu(p^\prime)$ is given as,
\begin{equation}
 M = \frac{G_F}{\sqrt{2}} [\bar u(k^\prime)\gamma^\mu (1-\gamma_5) v(k)][\bar u(p^\prime)\gamma_\mu (1-\gamma_5)u(p)]
\end{equation}
where $G_F$ is the Fermi constant. After squaring amplitude and solve it using trace technology, we get the spin averaged 
amplitude,
\begin{equation}
 \overline{|M|^2} = 64G_F^2 (p\cdot k)(p^\prime\cdot k^\prime)
 \label{amp:mu}
\end{equation}
The decay width of muon is,
\begin{align}
 d\Gamma &= \frac{d^3p^\prime}{(2\pi)^32E_{\nu_\mu}}\frac{d^3k^\prime}{(2\pi)^32E_e}\frac{d^3k}{(2\pi)^3 2E_{\bar \nu_e}}
 \frac{\overline{|M|^2}}{2E_\mu} (2\pi)^4\delta^4(p-p^\prime-k^\prime-k)
\end{align}
using the squared amplitude from eq.(\ref{amp:mu}), we get
\begin{align}
 d\Gamma &= \frac{32~G_F^2}{8(2\pi)^5 E_\mu}\frac{d^3k^\prime}{E_e}\frac{d^3p^\prime}{E_{\nu_\mu}}\frac{d^3k}{E_{\bar \nu_e}}
 \delta^4(p-p^\prime-k^\prime-k)(p\cdot k)(p^\prime \cdot k^\prime)
 \label{mu:dwa}
\end{align}
First we write eq.(\ref{mu:dwa}) as,
\begin{align}
 \Gamma &= \frac{32~G_F^2}{8(2\pi)^5 E_\mu} \int \frac{d^3k^\prime}{E_e} p^\alpha k^{\prime^\beta} I_{\alpha \beta}(p-k^\prime)
 \label{mu:dwb}
\end{align}
where
\begin{equation}
 I_{\alpha \beta}(p-k^\prime) \equiv \int \frac{d^3k}{E_{\bar \nu_e}}\frac{d^3p^\prime}{E_{\nu_\mu}}
 \delta^4 (p-p^\prime-k^\prime-k) k_\alpha p^\prime_\beta
\end{equation}
and then to find out $I_{\alpha \beta}(p-k^\prime)$, we use the generic phase space integral formula,
\begin{align}\nonumber
 I_{\alpha \beta} &\equiv \int \frac{d^3 p}{\sqrt{m_2^2+\vec p\cdot \vec p}} \frac{d^3 q}{\sqrt{m_1^2+\vec q \cdot \vec q}}
 \delta^4(k-p-q)p_\alpha q_\beta = \frac{I}{12 k^4}(k^2[k^2-(m_1-m_2)^2]\\
 & [k^2-(m_1+m_2)^2]g_{\alpha\beta}+2[k^4+k^2(m_1^2+m_2^2)-2(m_1^2-m_2^2)^2]k_\alpha k_\beta)
 \label{muon:psgf}
\end{align}
where
\begin{equation}
 I = \frac{2\pi}{k^2}\sqrt{[k^2-(m_1-m_2)^2][k^2-(m_1+m_2)^2]}.
\end{equation}
Applying this to our scenario by putting $m_1^2=m_{\bar \nu_e}^2 = \xi^\prime_3 k^3$, $m_2^2=m_{\nu_\mu}^2 = -\xi^\prime_3 {p^\prime}^3$ and taking $k=p^\prime/2 \sim p/4$, we find  
\begin{align}
 I_{\alpha \beta}(p-k^\prime)
 &= \frac{\pi}{6}\left[1+ \frac{7}{64}\frac{\xi^\prime_3 p^3}{(p-k^\prime)^2}\right]\\\nonumber
& \left([(p-k^\prime)^2+ \frac{7}{32}\xi^\prime_3 p^3]g_{\alpha\beta}
 + 2\left[1-\frac{7}{64}\frac{\xi^\prime_3 p^3}{(p-k^\prime)^2}\right](p-k^\prime)_\alpha (p-k^\prime)_\beta\right)
\end{align}
after contracting $I_{\alpha\beta}$ with the muon and electron momentums which respectively are $p$ and $k^\prime$, we get
\begin{align}
 p^\alpha k^{\prime^\beta}I_{\alpha \beta}(p-k^\prime) &= \frac{\pi}{6}\left[1+ \frac{7}{64}\frac{\xi^\prime_3 p^3}{(p-k^\prime)^2}\right]\\\nonumber
 &\left([(p-k^\prime)^2+\frac{7}{32}\xi^\prime_3 p^3](p \cdot k^\prime) + 2\left[1-\frac{7}{64}\frac{\xi^\prime_3 p^3}{(p-k^\prime)^2}\right]
 (p\cdot p-p\cdot k^\prime)(p\cdot k^\prime-k^\prime \cdot k^\prime)\right)
\end{align}
\begin{figure}[t!]
\vspace*{10 mm}
\begin{center}
\includegraphics[scale=1.20,angle=0]{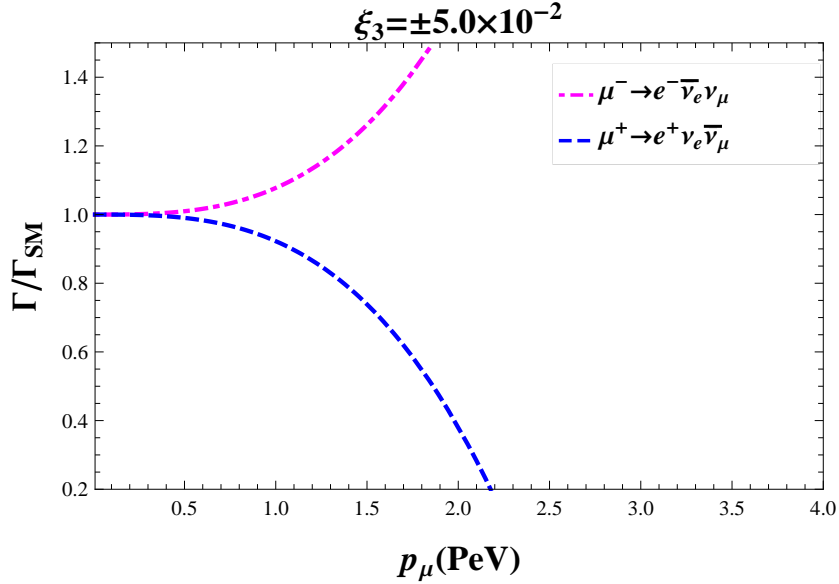}
\vspace*{3mm}
\caption{The ratio $\Gamma/\Gamma_{SM}$ for $\mu^+ \rightarrow e^+ \nu_e \bar \nu_\mu$ and $\mu^- \rightarrow e^- \bar\nu_e  \nu_\mu$ processes in Lorentz invariance violating framework to its standard model prediction for superluminal antineutrino $(\xi_3<0)$ and subluminal neutrino $(\xi_3>0)$ final states as a function of muon momentum $p_\mu$. Here we considered
$\xi_3=\pm 5.0\times 10^{-2}$.}
\label{fig:mu}
\end{center}
\end{figure} 
where,
\begin{align}\nonumber
 p\cdot p &= m^2_\mu\\\nonumber
 k^\prime \cdot k^\prime &= m_e^2 \approx 0\\\nonumber
 p\cdot k^\prime &= \vec k^\prime (E_\mu-\vec p \cos \theta)\\
 (p-k^\prime)^2 &= m^2_\mu - 2 \vec k^\prime (E_\mu-\vec p \cos \theta).
\end{align}
The decay width from eq.(\ref{mu:dwb}) can be written as,
\begin{align}
 \Gamma &= \frac{32 G_F^2}{8(2\pi)^5}\frac{(2\pi)}{E_\mu} \int^1_{-1} d\cos\theta \int^{m^2_\mu/2(E_\mu-k \cos\theta)}_0 
 k^{\prime} dk^\prime 
 p^\alpha k^{\prime^\beta}I_{\alpha \beta}
 \label{mu:dwe}
\end{align}
after solving it, we finally get,
\begin{equation}
 \Gamma = \frac{G^2_F m^4_\mu}{192\pi^3 E_\mu}\left(m^2_\mu + \frac{17}{80} \xi^\prime_3 p^3\right).
\end{equation}
We compare our result with the standard model prediction of muon decay in a moving frame, which is 
\begin{equation}
 \Gamma_{SM}(\mu\rightarrow e \bar \nu_e \nu_\mu) = \frac{G^2_F m^5_\mu}{192\pi^3}\frac{m_\mu}{E_\mu}.
\end{equation}
We compute the muon decay rate for subluminal neutrino $(\xi_3>0)$ and superluminal antineutrino $(\xi_3<0)$ 
and obtain the following:
\begin{itemize}
\item The decay rate of the process $\Gamma (\mu^-\rightarrow e^- \bar \nu_e \nu_\mu)$ is enhanced, as shown in Fig.(\ref{fig:mu}) for $\xi_3 = \pm 5.0\times 10^{-2}$.
\item The decay rate of the process $\Gamma (\mu^+\rightarrow e^+ \nu_e \bar\nu_\mu)$ is reduced, as shown in Fig.(\ref{fig:mu}) for $\xi_3 = \pm 5.0\times 10^{-2}$.
\end{itemize}
%
\subsection{$\boldmath {K^+\rightarrow \pi^0 e^+ \nu_e}$}
\begin{figure}[t!]
\vspace*{10 mm}
\begin{center}
\includegraphics[scale=1.10,angle=0]{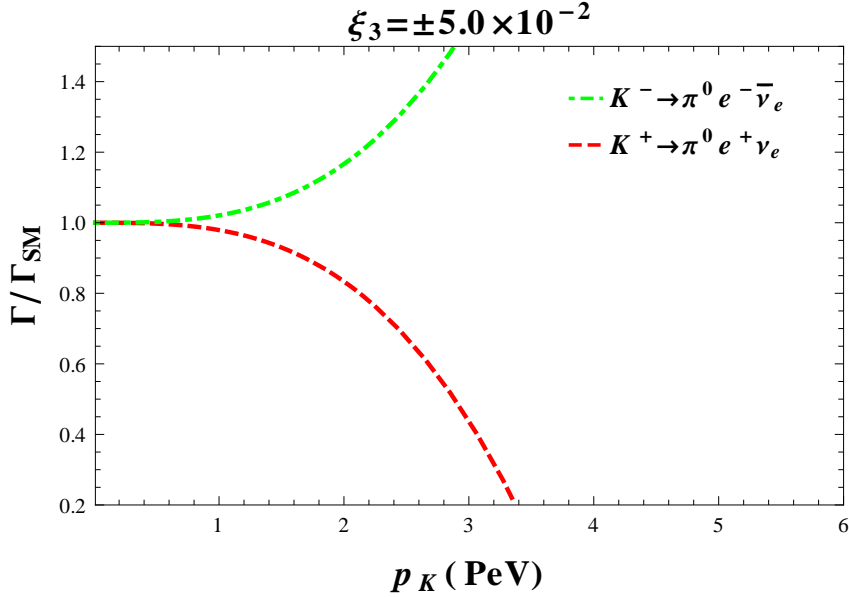}
\vspace*{3mm}
\caption{The ratio $\Gamma/\Gamma_{SM}$ for $K^+ \rightarrow \pi^0 e^+ \nu_e$ and $K^- \rightarrow \pi^0 e^- \bar\nu_e$ processes in Lorentz invariance violating framework to its standard model prediction for superluminal $\bar \nu_e~(\xi_3<0)$ and subluminal $\nu_e~(\xi_3>0)$ final states as a function of kaon momentum $p_K$. We considered $\xi_3= \pm 5.0\times 10^{-2}$ for corresponding processes.}
\label{fig:ka}
\end{center}
\end{figure}

We also calculate 3-body kaon decay width using the modified dispersion relation for neutrino by taking $n=3$ case. The 
amplitude calculation of kaon decay process $K^+(p_K) \rightarrow \pi^0(p_\pi) e^+(p_e) \nu_e(p_\nu)$ gives,
\begin{equation}
 \overline{|M|^2} = 16 G^2_F |V_{us}|^2 f^2_+ [m^2_K (p_K \cdot p_\nu + p_{\pi} \cdot p_\nu)-2(p_K \cdot p_\nu)
 (p_K \cdot p_\pi)-2 (p_K \cdot p_\nu)(p_K \cdot p_\nu)-m^2_K \xi^\prime_3 p^3_\nu]
\end{equation}
where $f_+$ is the kaon form factor. The Decay width of kaon is,
\begin{align}
 d\Gamma &= \frac{d^3p_{\pi}}{(2\pi)^3 2E_{\pi}}\frac{d^3p_{\nu_e}}{(2\pi)^3 2E_{\nu_e}}\frac{d^3p_e}{(2\pi)^3 2E_e}
 \frac{\overline{|M|^2}}{2E_K} (2\pi)^4\delta^4(p_K-p_{\pi}-p_{\nu_e}-p_e)
\end{align}
which gives,
\begin{align}
 \Gamma \simeq \frac{G^2_F |V_{us}|^2 f^2_+ m^4_K}{768 \pi^3 E_K} \left[m^2_K \left(1-\frac{8 m^2_\pi}{m^2_K}\right)
 -\frac{4}{9} p^3_K \xi^\prime_3 \left(1-\frac{m^4_\pi}{m^4_K}\right)\right]
 \label{ka:dwa}
\end{align}
It is clear from eq.(\ref{ka:dwa}) that the $K^+(K^-)$ decay rate goes down (up) as kaon momentum $p_K$ increases, which is 
shown in Fig.(\ref{fig:ka}) for $\xi_3=\pm5.0 \times 10^{-2}$.
%
\subsection{$\boldmath {n \rightarrow p^+ e^- \bar \nu_e}$}

In the similar way like muon decay, we also calculate the neutron beta decay width using the modified dispersion relation
for antineutrino. The spin averaged amplitude squared for the neutron decay process 
$n(p) \rightarrow p^+(k) e^-(k^\prime)\bar \nu_e(p^\prime)$ comes,
\begin{equation}
  \overline{|M|^2} = 64 G^2_F (p \cdot p^\prime)(k \cdot k^\prime)
  \label{ams:nutron}
\end{equation}
using eq.(\ref{ams:nutron}), we get the following decay width of neutron,
\begin{align}
  d \Gamma = \frac{32~G^2_F}{8(2\pi)^5 E_n}\frac{d^3k}{E_p}\frac{d^3k^\prime}{E_e}\frac{d^3p^\prime}{E_{\bar \nu_e}}
  \delta^4(p-k-k^\prime-p^\prime)(p \cdot p^\prime)(k \cdot k^\prime)
  \label{neutron:dw}
\end{align}
we solve eq.(\ref{neutron:dw}) in the similar way like muon decay using generic phase space integral formula 
(eq.\ref{muon:psgf}). Then we solve the final integral over the electron energy, for which the minimum energy is the rest
energy $m_e$ of the electron while the maximum energy is approximately,
\begin{equation}
 E_{max} \approx m_n - m_p
\end{equation}
which finally gives,
\begin{equation}
 \Gamma \sim \frac{G^2_F (m_n-m_p)^3 m_n}{15\pi^3 E_n}\left[(m_n-m_p)^2 -\frac{5}{16} \xi^\prime_3 p^3\right] 
 \label{nutron:dwa}
\end{equation}
For $\xi_3=0.05$ the neutron decay width goes down at neutrino momentum $p\simeq 0.1$ PeV. 
This implies that antineutrino production from neutron decay will be suppressed and so in our model, it is also possible to explain the absence of Glashow resonance \cite{Glashow:1960zz}. The decay rate of the charge conjugate process $\bar n \rightarrow \bar p e^+ \nu_e$ is enhanced, but since only neutrons are produced in the $p+\gamma \rightarrow \Delta \rightarrow n+ \pi^+$ processes at the source, the enhanced decay of $\bar n$ is not relevant to the IceCube events. 
\section{Conclusion}\label{sec:conclusion}
In this paper we provide a mechanism by which one can account for the lack of antineutrino events at Glashow resonance (6.3 PeV) at IceCube. We show that if the neutrino (antineurino) dispersion is modified by leading order Planck scale suppression 
$E^2= p^2 - (\xi_3/M_{Pl}) p^3$ (where $\xi_3>0$ correspond to neutrinos and $\xi_3<0$ correspond to 
antineutrino), then there is a suppression of the $\pi^+$ decay width and corresponding neutrinos will be cutoff at energies $E_\nu =1.3$ PeV (with $\xi_3=0.05$). The neutrinos from Kaon decay $K^+\rightarrow \mu^+ \nu_\mu$ will be cutoff at 2 PeV. 
\begin{itemize}
 \item Three body decays like $\mu^-\rightarrow e^- \bar \nu_e \nu_\mu$ and 
 $K^- \rightarrow \pi^0 e^- \bar \nu_e$ get enhanced due to different $\xi_3$ dependence in their $\overline{|M|^2}$, whereas 
three body decay widths of $\mu^+$ and $K^+$ get suppressed.
 \item Neutron decay $n \rightarrow p^+ e^- \bar \nu_e$ gets suppressed in the similar way as $\mu^+$ decay. So
 if the source of $\bar\nu_e$ is neutron beta-decay then the mechanism proposed in this paper can be used to explain the 
 absence of Glashow resonance at IceCube.
 \item Radiative three body decays like $\pi^\pm \rightarrow e^\pm \nu \gamma$ and $\pi^\pm \rightarrow \mu^\pm \nu \gamma$ are factorized to the $\overline{|M|^2}$ for two body decays $\pi^\pm \rightarrow e^\pm \nu$ and $\pi^\pm \rightarrow \mu^\pm \nu$ times $\alpha_{em}$ 
\cite{Agashe:2014kda,Bijnens:1992en} and these are also suppressed like two body decay processes.
\end{itemize}
The enhancement in $\mu^-$ decay will be significant at muon energies of 2 PeV and if the primary source of $\mu^-$ is $\pi^-$ decay then there will be no observable consequence of this in IceCube events. 
However such enhancement of the $\mu^-$ decay rate would be observable for $\mu^-$ produced not from $\pi^-$ decay but e.g via pair production e.g in $e^+ e^- \rightarrow \mu^+ \mu^-$. The precise numerical values depend on the choice
of the parameter $\xi_3$, but obviously a cutoff between $\sim 3$ PeV and 6.3 PeV can be easily obtained in this model. We conclude that if neutrinos at Glashow resonance energies are not observed at IceCube then explanations in terms of new physics such as Lorentz violating modified neutrino dispersion relation become attractive. 
The fact that neutron decay into $p+e+\bar \nu_e$ is suppressed has the following implications. The conventional 
$\pi/K$ decay neutrinos from astrophysical sources have cutoff in the range of $\sim 3$ PeV. However the B-Z neutrinos which arise in GZK process have two components \cite{Engel:2001hd}, the higher energy neutrinos from $\pi/K$ will be more suppressed compared to the lower energy $n$ decay to $\bar\nu_e$. But both components of GZK process will be suppressed at $E_\nu > 3$ PeV.
\section{Acknowledgement}
One of us (S.P) would like to thank  Prof. Subhendra Mohanty and PRL for hospitality and support while this work was 
started, and to acknowledge the hospitality of Prof. Georg Raffelt and MPI, Munich while this work was continued. S.M would
like to thank Alan Kostelecky for valuable discussion.
\appendix
\section{Dispersion Relation}
\label{ap:dspr}
The cubic dispersion relation we used for neutrinos and antineutrinos can be obtained from the dimension 5 operator 
\cite{Myers:2003fd,Diaz:2013wia},
\begin{equation}\label{lfdr}
 {\cal L}_{LV} = \frac{1}{M_{pl}} \bar \psi (\eta_1 \slashed n + \eta_2 \slashed n \gamma_5)(n \cdot \partial)^2 \psi
\end{equation}
where $n_\mu$ is a fixed four vector that specifies the preferred frame. Both the vector and axial-vector terms
in eq.(\ref{lfdr}) are CPT violating in addition to being Lorentz violating. The Lagrangian gives the equation of 
motion,
\begin{equation}
 i \slashed \partial \psi = -\frac{1}{M_{pl}} (\eta_1 \slashed n + \eta_2 \slashed n \gamma_5)(n \cdot \partial)^2 \psi
\end{equation}
where we have taken $E\gg m$. This leads to the following dispersion relation for left and right handed particles $\psi$,
\begin{equation}
 E^2 = p^2  + 2 (\eta_1 \pm \eta_2) \frac{p^3}{M_{pl}} 
\end{equation}
where $+$ and $-$ signs correspond to $\psi_R$ and $\psi_L$ respectively. Now taking the charge conjugation of 
eq.(\ref{lfdr}), we find
\begin{equation}\label{lafdr}
 {\cal L}_{LV} = \frac{1}{M_{\tiny\mbox{pl}}} \bar \psi^c (-\eta_1 \slashed n + \eta_2 \slashed n \gamma_5)(n \cdot \partial)^2 \psi^c
\end{equation}
where we used charge conjugation properties viz. $C^{-1}\gamma_\mu C=-\gamma_\mu$ and 
$C^{-1}\gamma_\mu\gamma_5 C=\gamma_\mu\gamma_5$. The operator (eq.\ref{lafdr}) gives the following dispersion relation for left
and right handed antiparticle $\psi^c$,
\begin{equation}
 E^2 = p^2  + 2(-\eta_1 \pm \eta_2) \frac{ p^3}{M_{pl}} 
\end{equation}
%
%
where the $+$ sign is for $\psi^c_R$ and $-$ sign is for $\psi^c_L$. Therefor for the case of left-handed neutrinos $\nu_L$,
we will have the dispersion relation,
\begin{equation}
 E^2 = p^2 + 2(\eta_1-\eta_2) \frac{p^3}{M_{pl}}
\end{equation}
and for antineutrinos $\nu^c_R$ we have,
\begin{equation}
 E^2 = p^2 - 2(\eta_1-\eta_2) \frac{p^3}{M_{pl}}
\end{equation}
We have dispersion relation for neutrinos and antineutrinos $E^2 = p^2-(\xi_3/M_{pl})p^3$, where $\xi_3 = -2(\eta_1-\eta_2)$ for neutrinos and $\xi_3 = 2(\eta_1-\eta_2)$ for antineutrinos.
\section{Spinors Relation}
We assume that all the particles expect neutrinos follow the standard 
energy-momentum relation i.e,
\begin{equation}
 E_i = \sqrt{p_i^2 + m_i^2},
\end{equation}
where $m_i$ and $p_i$ are the mass and momentum of different particles $(i=e,\mu,\tau~etc)$. The neutrinos follow the modified 
dispersion relation given in eq.$(\ref{dr})$. 
There exist very stringent bounds \cite{Giudice:2011mm}, which suggest that neutrino flavor is 
independent of their dispersion relation, so we assumed the universal dispersion relation for different flavor of neutrinos. 
We also define,
\begin{equation}
 F(p) \equiv \frac{E}{p} = 1-\frac{\xi_n p^{n-2}}{2 M_{pl}^{n-2}},
\end{equation}
where the function $F(p)$ is the measure of the deviation of neutrino dispersion relation from the standard 
one \cite{Mannarelli:2011tv}. In this framework, the modified Dirac equation for neutrino can be written as,
\begin{equation}
 (i\gamma^0 \partial_0 - i F(p) \vec \gamma \cdot \vec \partial)\psi(x)=0
\end{equation}
where we have neglected the neutrino mass for simplification. Now we replace the Dirac field $\psi$ in terms of the linear 
combination of plane waves i.e,
\begin{equation}
 \psi(x) = u(p) e^{-ip\cdot x}
\end{equation}
using it, we get the following form of Dirac equation,
\begin{equation}
 (\gamma^0 E - F(p) \vec \gamma \cdot \vec p)u(p)=0.
 \label{ap:de}
\end{equation}
Clearly, the positive energy solution of this equation will satisfy,
\begin{equation}
 E(p) = F(p)p,
 \label{mpr}
\end{equation}
we used these results in the derivation of the spinors sum of neutrinos, which comes,
\begin{equation}
 \sum_{s=1,2} u^s(p) \bar u^s(p) = \begin{pmatrix}
                                    0 & \tilde p\cdot \sigma\\
                                    \tilde p \cdot \bar\sigma & 0
                                   \end{pmatrix}
\end{equation}
where we assumed neutrino to be massless and defined $\tilde p = (E, F(p)p)$. Following the Dirac algebra, we get 
the following result for spinor sum,
\begin{equation}
  \sum_{s=1,2} u^s(p) \bar u^s(p) = \gamma^\mu \tilde p_\mu \equiv F(p) \gamma^\mu p_\mu
  \label{sr}
\end{equation}
where we used the result of eq.(\ref{mpr}) for further simplification.
For antiparticle when $m=0$, there is an overall negative sign in eq.(\ref{ap:de}) and following the same procedure we obtain
the same result,
\begin{equation}
  \sum_{s=1,2} v^s(p) \bar v^s(p) = F(p) \gamma^\mu p_\mu
  \label{sra}
\end{equation}
%

\end{document}